\pdfoutput=1
\documentclass[12pt]{article}
\usepackage{lineno}  % for line numbering during review

\usepackage{graphicx}  % to include figures (can also use other packages)
\usepackage{xspace}
\usepackage{color}
\usepackage{colortbl}
\usepackage{subfigure}
\usepackage{amsmath}

\usepackage{ifthen} % for conditional statements
\newboolean{pdflatex}
\setboolean{pdflatex}{true} % use this if using non-eps figures
\usepackage{rotating} 
\newboolean{articletitles}
\setboolean{articletitles}{true} 

\newboolean{uprightparticles}
\setboolean{uprightparticles}{false} %Set to true to get roman particle symbols
\usepackage{amssymb}
\usepackage{amsfonts}
\usepackage{upgreek}
\ifthenelse{\boolean{uprightparticles}}%
{

 \def\PDelta      {\ensuremath{\Delta}\xspace}                 
 \def\PXi      {\ensuremath{\Xi}\xspace}                 
 \def\PLambda      {\ensuremath{\Lambda}\xspace}                 
 \def\PSigma      {\ensuremath{\Sigma}\xspace}                 
 \def\POmega      {\ensuremath{\Omega}\xspace}                 
 \def\PUpsilon      {\ensuremath{\Upsilon}\xspace}                 
 
 \def\PB      {\ensuremath{\mathrm{B}}\xspace}                 
                  
 \def\PD      {\ensuremath{\mathrm{D}}\xspace}

 \def\PK      {\ensuremath{\mathrm{K}}\xspace}

 \def\Pc      {\ensuremath{\mathrm{c}}\xspace}

 \def\Pi      {\ensuremath{\mathrm{i}}\xspace}

}
{

 \mathchardef\PDelta="7101
 \mathchardef\PXi="7104
 \mathchardef\PLambda="7103
 \mathchardef\PSigma="7106
 \mathchardef\POmega="710A
 \mathchardef\PUpsilon="7107
                  
 \def\PB      {\ensuremath{B}\xspace}                 
                  
 \def\PD      {\ensuremath{D}\xspace}

 \def\PK      {\ensuremath{K}\xspace}

 \def\Pc      {\ensuremath{c}\xspace}

 \def\Pi      {\ensuremath{i}\xspace}

}

\def\c     {\ensuremath{\Pc}\xspace}

\def\kaon  {\ensuremath{\PK}\xspace}
  \def\Kbar  {\kern 0.2em\overline{\kern -0.2em \PK}{}\xspace}

\def\Kz    {\ensuremath{\kaon^0}\xspace}
\def\Kzb   {\ensuremath{\Kbar^0}\xspace}
\def\KzKzb {\ensuremath{\Kz \kern -0.16em \Kzb}\xspace}
\def\Kp    {\ensuremath{\kaon^+}\xspace}
\def\Km    {\ensuremath{\kaon^-}\xspace}

\def\KpKm  {\ensuremath{\Kp \kern -0.16em \Km}\xspace}

  \def\Dbar    {\kern 0.2em\overline{\kern -0.2em \PD}{}\xspace}
\def\D       {\ensuremath{\PD}\xspace}

\def\Dz      {\ensuremath{\D^0}\xspace}
\def\Dzb     {\ensuremath{\Dbar^0}\xspace}
\def\DzDzb   {\ensuremath{\Dz {\kern -0.16em \Dzb}}\xspace}
\def\Dp      {\ensuremath{\D^+}\xspace}
\def\Dm      {\ensuremath{\D^-}\xspace}

\def\DpDm    {\ensuremath{\Dp {\kern -0.16em \Dm}}\xspace}

  \def\Bbar    {\kern 0.18em\overline{\kern -0.18em \PB}{}\xspace}

\def\Bs      {\ensuremath{\B^0_s}\xspace}
\def\Bsb     {\ensuremath{\Bbar^0_s}\xspace}

  \def\Y#1S{\ensuremath{\PUpsilon{(#1S)}}\xspace}% no space before {...}!

\newcommand{\tev}{\ensuremath{\mathrm{\,Te\kern -0.1em V}}\xspace}
\newcommand{\gev}{\ensuremath{\mathrm{\,Ge\kern -0.1em V}}\xspace}
\newcommand{\mev}{\ensuremath{\mathrm{\,Me\kern -0.1em V}}\xspace}
\newcommand{\kev}{\ensuremath{\mathrm{\,ke\kern -0.1em V}}\xspace}
\newcommand{\ev}{\ensuremath{\mathrm{\,e\kern -0.1em V}}\xspace}
\newcommand{\gevc}{\ensuremath{{\mathrm{\,Ge\kern -0.1em V\!/}c}}\xspace}
\newcommand{\mevc}{\ensuremath{{\mathrm{\,Me\kern -0.1em V\!/}c}}\xspace}
\newcommand{\gevcc}{\ensuremath{{\mathrm{\,Ge\kern -0.1em V\!/}c^2}}\xspace}
\newcommand{\gevgevcccc}{\ensuremath{{\mathrm{\,Ge\kern -0.1em V^2\!/}c^4}}\xspace}
\newcommand{\mevcc}{\ensuremath{{\mathrm{\,Me\kern -0.1em V\!/}c^2}}\xspace}

\def\to                 {\ensuremath{\rightarrow}\xspace}

\def\CP                {\ensuremath{C\!P}\xspace}

\def\gsim{{~\raise.15em\hbox{$>$}\kern-.85em
          \lower.35em\hbox{$\sim$}~}\xspace}
\def\lsim{{~\raise.15em\hbox{$<$}\kern-.85em
          \lower.35em\hbox{$\sim$}~}\xspace}

\def\pythia     {\mbox{\textsc{Pythia}}\xspace}

\def\AT#1     {\ensuremath{A_T^{#1}}\xspace}           % 2

\def\C#1      {\ensuremath{\mathcal{C}_{#1}}}                       % 9
\def\Cp#1     {\ensuremath{\mathcal{C}_{#1}^{'}}}                    % 7
\def\Ceff#1   {\ensuremath{\mathcal{C}_{#1}^{\mathrm{(eff)}}}}        % 9  
\def\Cpeff#1  {\ensuremath{\mathcal{C}_{#1}^{'\mathrm{(eff)}}}}       % 7
\def\Ope#1    {\ensuremath{\mathcal{O}_{#1}}}                       % 2
\def\Opep#1   {\ensuremath{\mathcal{O}_{#1}^{'}}}                    % 7

\usepackage{hyperref}
\usepackage[all]{hypcap} 
\usepackage{bm}
\usepackage{afterpage}

\renewcommand{\Bs}{\overline{B}_s^0}

\newcommand{\bea}{\begin{eqnarray}}
\newcommand{\eea}{\end{eqnarray}}
\newcommand{\beq}{\begin{equation}}
\newcommand{\eeq}{\end{equation}}

\usepackage{mciteplus}

\begin{document}
\renewcommand{\thefootnote}{\fnsymbol{footnote}}
\setcounter{footnote}{1}

%%%%%%%%%%%%%%%%%%%%%%%%%
%%%%% Title     %%%%%%%%%
%%%%%%%%%%%%%%%%%%%%%%%%%

% %%%%%%% CHOOSE --------
%  Choose the right title template or customize existing one if your type is still missing

% $Id: titlepage.tex 2394 2011-02-17 10:11:10Z gcowan $
% ===============================================================================
% Purpose: LHCb-ANA Note title page template
% Author: 
% Created on: 2010-10-05
% ===============================================================================

%%%%%%%%%%%%%%%%%%%%%%%%%
%%%%  TITLE PAGE  %%%%%%
%%%%%%%%%%%%%%%%%%%%%%%%%
\begin{titlepage}
%Primary authors Bilas Pal, Sheldon Stone and Liming Zhang
% Header ---------------------------------------------------
\belowpdfbookmark{Title page}{title}

\pagenumbering{roman}
\vspace*{-1.5cm}
\centerline{\large EUROPEAN ORGANIZATION FOR NUCLEAR RESEARCH (CERN)}
\vspace*{1.5cm}
\hspace*{-5mm}\begin{tabular*}{16cm}{lc@{\extracolsep{\fill}}r}
%\vspace*{-12mm}\mbox{\!\!\!\includegraphics[width=.12\textwidth]{lhcb-logo}}& & \\
&&CERN-PH-EP-2011-212\\
&& LHCb-PAPER-2011-026\\
&&December 20, 2011 \\
\end{tabular*}
\vspace*{4cm}
\begin{center}

{\bf\huge\boldmath 
Observation of $\Bs\to J/\psi f'_2(1525)$ in $J/\psi K^+K^-$ final states}\\
\vspace*{2cm}
\normalsize {
The LHCb Collaboration\footnote{Authors are listed on the following pages.}
%========================================================================%
}
\end{center}

\vspace*{2.0cm}

\vspace*{-2cm}
% Abstract -----------------------------------------------
\begin{abstract}
  \noindent
%Measurement of mixing induced \CP violation in $\Bs$ decays is of prime importance in probing new physics. 
 The decay $\Bs\to J/\psi K^+ K^-$ is investigated using 0.16\,fb$^{-1}$ of data collected with the LHCb detector using 7 TeV $pp$ collisions. Although the $J/\psi \phi$ channel is well known, final states at higher $K^+K^-$ masses have not previously been studied. In the  $K^+K^-$ mass spectrum we observe a significant signal in the
$f'_2(1525)$ region as well as a non-resonant component. After subtracting the non-resonant component, we find ${{\cal{B}}\left(\Bs\to J/\psi f'_2(1525)\right)}/{{\cal{B}}\left(\Bs\to J/\psi \phi\right)}=(26.4\pm2.7\pm2.4)$\%.
\end{abstract}

\vspace*{2.0cm}
\hspace*{6mm}{\it PACS:} 13.25.Hw, 14.40.Nd, 14.40.Be\\
\hspace*{6mm}Submitted to Physical Review Letters\\

\newpage
% Authors -------------------------------------------------
\begin{center}
The LHCb Collaboration
\begin{flushleft}
R.~Aaij$^{23}$, 
C.~Abellan~Beteta$^{35,n}$, 
B.~Adeva$^{36}$, 
M.~Adinolfi$^{42}$, 
C.~Adrover$^{6}$, 
A.~Affolder$^{48}$, 
Z.~Ajaltouni$^{5}$, 
J.~Albrecht$^{37}$, 
F.~Alessio$^{37}$, 
M.~Alexander$^{47}$
% ...

{\footnotesize \it
$ ^{1}$Centro Brasileiro de Pesquisas F\'{i}sicas (CBPF), Rio de Janeiro, Brazil\\
% ...
$ ^{5}$Clermont Universit\'{e}, Universit\'{e} Blaise Pascal, CNRS/IN2P3, LPC, Clermont-Ferrand, France\\
$ ^{6}$CPPM, Aix-Marseille Universit\'{e}, CNRS/IN2P3, Marseille, France\\
% ...
$ ^{23}$Nikhef National Institute for Subatomic Physics, Amsterdam, The Netherlands\\
% ...
$ ^{35}$Universitat de Barcelona, Barcelona, Spain\\
$ ^{36}$Universidad de Santiago de Compostela, Santiago de Compostela, Spain\\
$ ^{37}$European Organization for Nuclear Research (CERN), Geneva, Switzerland\\
% ...
$ ^{42}$H.H. Wills Physics Laboratory, University of Bristol, Bristol, United Kingdom\\
% ...
$ ^{47}$School of Physics and Astronomy, University of Glasgow, Glasgow, United Kingdom\\
$ ^{48}$Oliver Lodge Laboratory, University of Liverpool, Liverpool, United Kingdom\\
% ...
$ ^{n}$LIFAELS, La Salle, Universitat Ramon Llull, Barcelona, Spain\\
% ...
}
\end{flushleft}
\end{center}

%%\vspace*{2.0cm}
%%\vspace{\fill}

\end{titlepage}
\renewcommand{\thefootnote}{\arabic{footnote}}
\setcounter{footnote}{0}
%\tableofcontents
\newpage

\pagestyle{empty}  % no page number for the title 

%%%%%%%%%%%%%%%%%%%%%%%%%%%%%%%%
%%%%%  EOD OF TITLE PAGE  %%%%%%
%%%%%%%%%%%%%%%%%%%%%%%%%%%%%%%%

%  empty page follows the title page ----

%\newpage
%\setcounter{page}{2}
%\mbox{~}
%

%\input{title-LHCb-CONF}
%\input{title-LHCb-PAPER}
% %%%%%%%%%%%%% ---------

%%%%%%%%%%%%%%%%%%%%%%%%%%%%%%%%
%%%%%  Table of Content   %%%%%%
%%%%%%%%%%%%%%%%%%%%%%%%%%%%%%%%
%%%% Uncomment next 2 lines if desired

%\tableofcontents

%%%%%%%%%%%%%%%%%%%%%%%%%
%%%%% Main text %%%%%%%%%
%%%%%%%%%%%%%%%%%%%%%%%%%

\pagestyle{plain} % restore page numbers for the main text
\setcounter{page}{1}
\pagenumbering{arabic}

% %%%%%%% CHOOSE --------
%% ----------------------------------
%% Line numbering on the left margin 
%% ----------------------------------
%% Uncomment during review phase. 
%% Comment it out before a final submission.

%% --------------------------------
% %%%%%%%%%%%%% ---------

%  You can include short sections directly in the main tex file. 
%  However, for larger papers it is desirable to split 
%  the text into several semiautonomous files, which can be revised independently. 
%  This is especially useful when developing a document in collaboration with several people, 
%  since then different parts can be edited independently. 
%  This type of file organization is shown here. 
% 

%\clearpage
%{\noindent\bf\Large Appendix}
%\appendix
%\input{appendixA}

%\addcontentsline{toc}{section}{References}

%\section{Introduction}

The $\Bs\to J/\psi K^+K^-$ channel has previously been studied only when the $K^+K^-$ are consistent with the decay of the $\phi$ meson. This mode has been used to measure the 
\CP violation in $\Bs$ mixing, $\phi_s$, a key probe in the search for physics beyond the Standard Model \cite{LHCb:2011ab,JpsiphiLHCb,Abazov:2011ry,Aaltonen:2011cq}.\footnote{Charge conjugate modes are also considered throughout.} In addition to the $\phi$ other resonant or non-resonant final states may appear and affect the \CP measurements, including an S-wave contribution \cite{Stone:2008ak}. In this paper we study the entire $K^+K^-$ mass spectrum, including a search for other final states that may be useful in \CP violation studies. These states may provide other ways of measuring $\phi_s$, in decays with a different spin structure that may be useful for revealing different aspects of \CP violation.

%\section{Data sample and analysis requirements}

We use a 0.16\,fb$^{-1}$  data sample collected with the LHCb detector  \cite{LHCb-det} at a center-of-mass energy of 7 TeV. The detector elements most important for this analysis include a vertex locator,  a silicon strip device that surrounds the $pp$ interaction region in the LHC, and other downstream tracking devices before and after a 4\,Tm dipole magnet. Two ring-imaging Cherenkov detectors are used to identify charged hadrons, while muons are identified using their penetration through iron. 
%Chambers interspersed with iron are used to distinguish muons from hadrons.
% The detector elements are placed along the beam line of the LHC starting with the Vertex Locator (VELO), a silicon strip device that surrounds the proton-proton interaction region and is positioned 8 mm from the beam during collisions. It provides precise locations for primary $pp$ interaction vertices, the locations of decays of long-lived particles, and contributes to the measurement of track momenta.  Other devices used to measure track momenta comprise a large area silicon strip detector (TT) located in front of a 3.7 Tm dipole magnet, and a combination of
%silicon strip detectors (IT) and straw drift chambers (OT) placed behind. Two Ring Imaging Cherenkov (RICH) detectors are used to identify charged hadrons. Further downstream an Electromagnetic Calorimeter (ECAL) is used for photon detection and electron identification, followed by a Hadron Calorimeter (HCAL), and  a system consisting of alternating layers of iron and chambers (MWPC and triple-GEM) that distinguishes muons from hadrons (MUON). The ECAL, MUON, and HCAL provide the capability of first-level hardware triggering.
This analysis is restricted to events accepted by a di-muon trigger \cite{Aaij:2011jh}.
Subsequent selection criteria are applied that serve to reject background, yet preserve high efficiencies, as determined by Monte Carlo (MC) events generated using \pythia \cite{Sjostrand:2006za}, and the LHCb detector simulation based on G{\sc eant} \cite{Agostinelli:2002hh}. 
To be considered as a $J/\psi\to\mu^+\mu^-$ candidate, opposite sign tracks are required to have  transverse momentum, $p_{\rm T}$, greater than 500 MeV, be identified as muons, and give a good fit to a
common  vertex.\footnote{We work in units where $c=1$.}  Di-muon candidates with masses between $-48$ and +43 MeV of the $J/\psi$ peak are selected for further analysis, where the r.m.s. resolution is 13.4 MeV. The invariant mass of the $\mu^+\mu^-$ pair is constrained to the $J/\psi$ mass for further analysis. 

Kaon candidates are selected if their minimum distance from the closest primary vertex is inconsistent with being produced at that vertex. They must be positively identified
based on the logarithm of the likelihood ratio comparing two particle hypotheses (DLL). There are two criteria used; loose corresponds to DLL($K-\pi)>0$, while tight has DLL($K-\pi)>10$ and DLL($K-p)>-3$. We use the loose criterion for checking kaon identification efficiencies, otherwise the tight criterion is used. 
In addition, the two kaons must have the sum of the magnitudes of their $p_{\rm T}> 900$\,MeV.

To select $\Bs$ candidates we require that the $K^+K^-$ pair and the $J/\psi$ candidate give a good fit to a common secondary vertex with a $\chi^2<5$ per degree of freedom. We also require that the $\Bs$ candidate's decay point must be  $>$ 1.5\,mm from the primary vertex and that the negative of its momentum vector points back to the primary.
%To select $\Bs$ candidates we further require that the two kaons vertex with a $\chi^2< 10$, that they form a candidate $\Bs$ vertex with the $J/\psi$ where the vertex fit $\chi^2$/ndof $<5$, and that this $\Bs$ candidate points to the primary vertex at an angle not different from its momentum direction by more than 0.68$^{\circ}$, with the impact parameter $\chi^2$ of the $\Bs$  less than 25. In addition, the $\Bs$ candidate must be inconsistent with decaying near the primary vertex with a flight distance larger than 1.5 mm.

\begin{figure}[hbt]
\centering
\includegraphics[width=4.in]{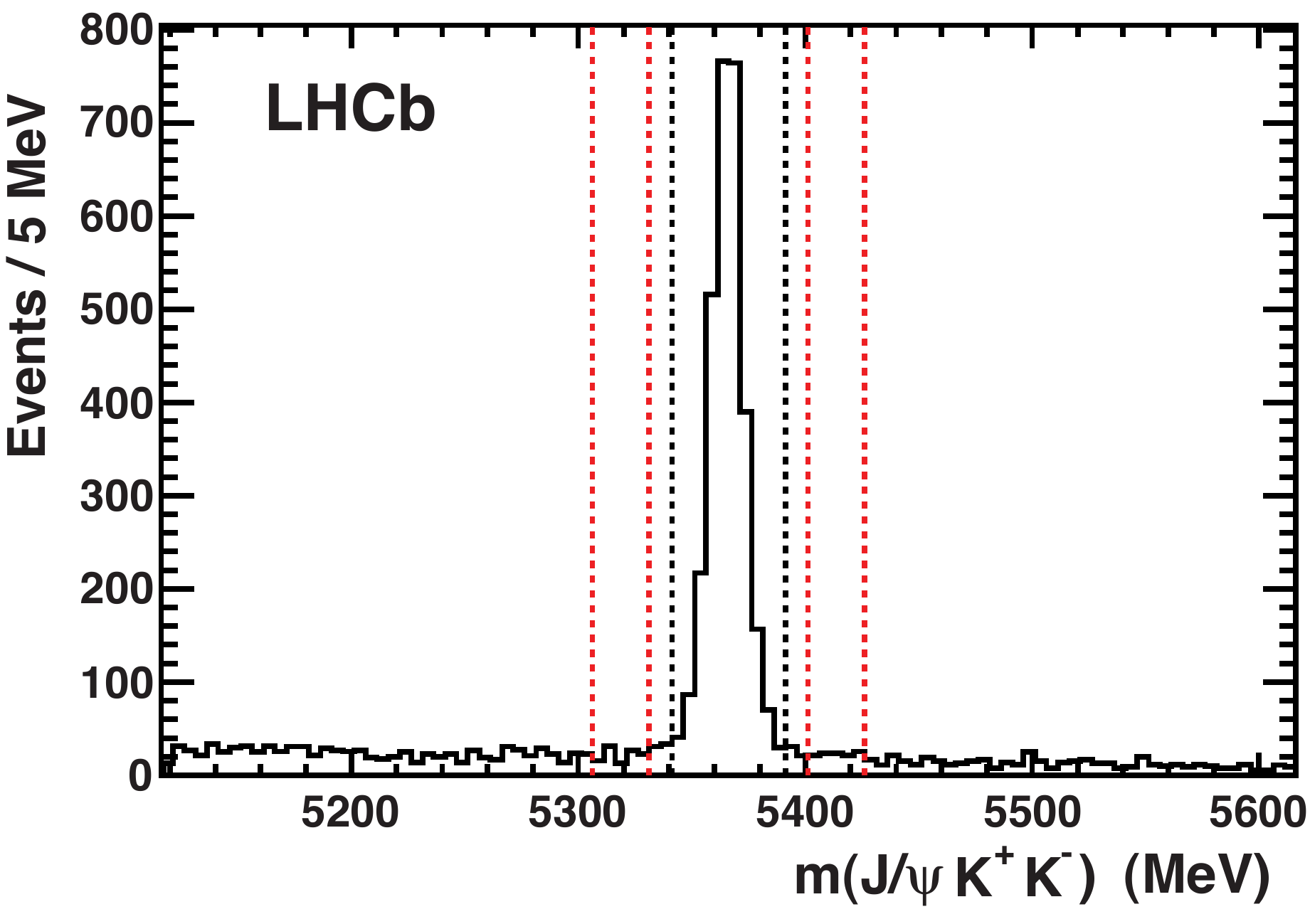}
\caption{Invariant mass of $J/\psi K^+K^-$ combinations. The vertical
lines indicate the signal and sideband regions. 
 } \label{mBsKK}
\end{figure}

\begin{figure}[hbt]
\centering
\includegraphics[width=4.in]{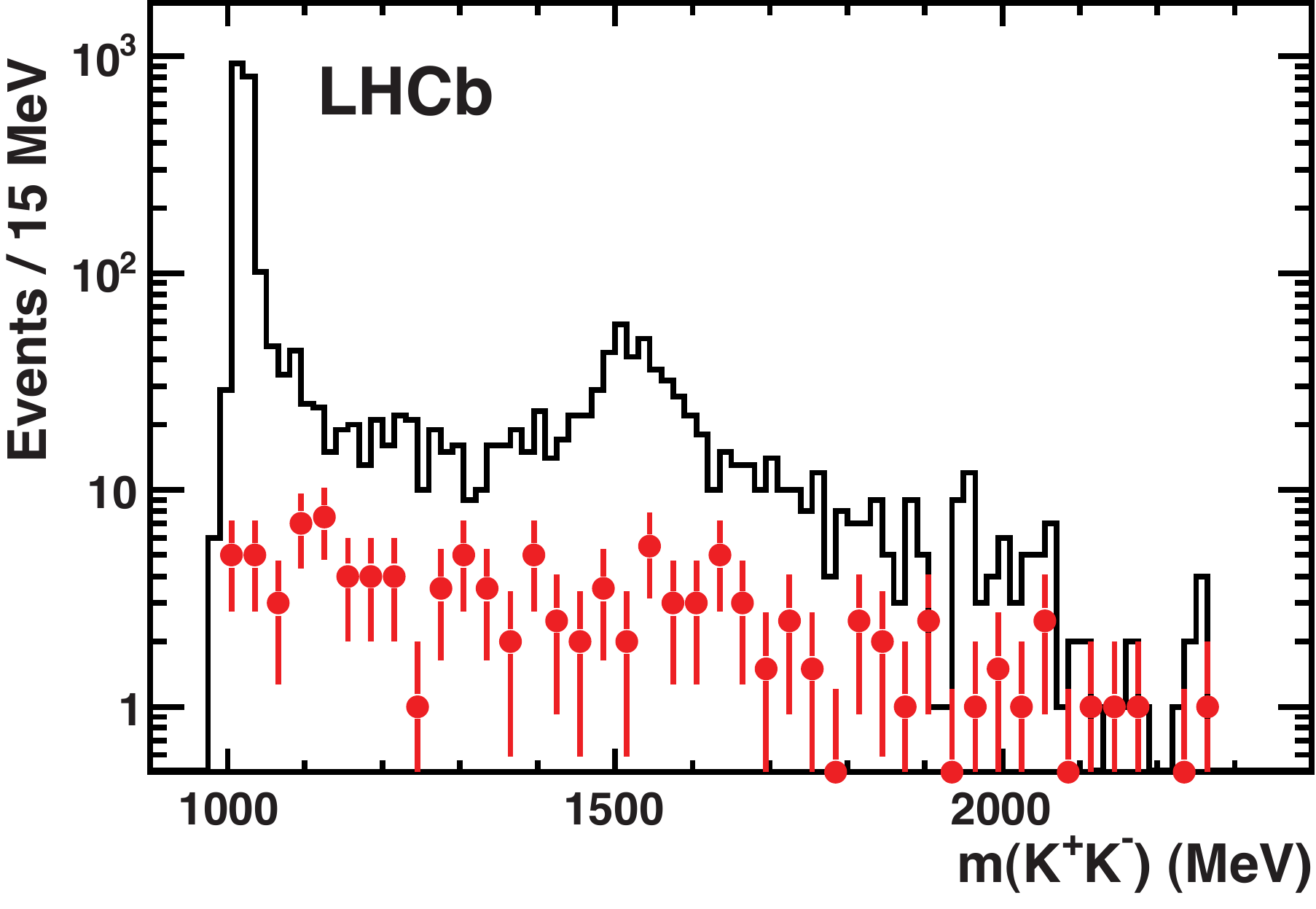}
\caption{Invariant mass of $K^+K^-$ combinations. The histogram shows the data in the signal region while the points (red) show the sidebands. } \label{mKK}
\end{figure}

%\subsection{Analysis of $K^+K^-$}
The  $\Bs$ candidate invariant mass is shown in Fig.~\ref{mBsKK}.  A clear signal is seen, part of which comes from the previously known $J/\psi\phi$ mode.
A check was made for possible resonant states decaying to $J/\psi K^-$ since similar exotic states have been claimed \cite{Z4430}, but no obvious structures are visible in the invariant mass spectrum. Figure~\ref{mKK} shows the $K^+K^-$ invariant mass for both signal and sideband regions, where the signal region extends $\pm$25 MeV around the $\Bs$ mass peak and the sidebands extend from 35 MeV to 60 MeV on either side of the peak.
Apart from the large peak at the $\phi$ there is a structure near 1525 MeV. In addition there is an excess of signal events over most of the mass range which we refer to as non-resonant. 
We investigate the possibility of the peak to be the $f'_2(1525)$ resonance. The PDG quotes the mass of the $f_2'$ state as 1525$\pm$5 MeV and the width as 73$^{+6}_{-5}$ MeV \cite{PDG}. Other states such as the $f_2(1270)$ and the $f_0(1500)$ have small branching fractions into $K^+K^-$ of less than 5\%, and are unlikely to have large rates.

\begin{figure}[hbt]
\centering
\includegraphics[width=6.in]{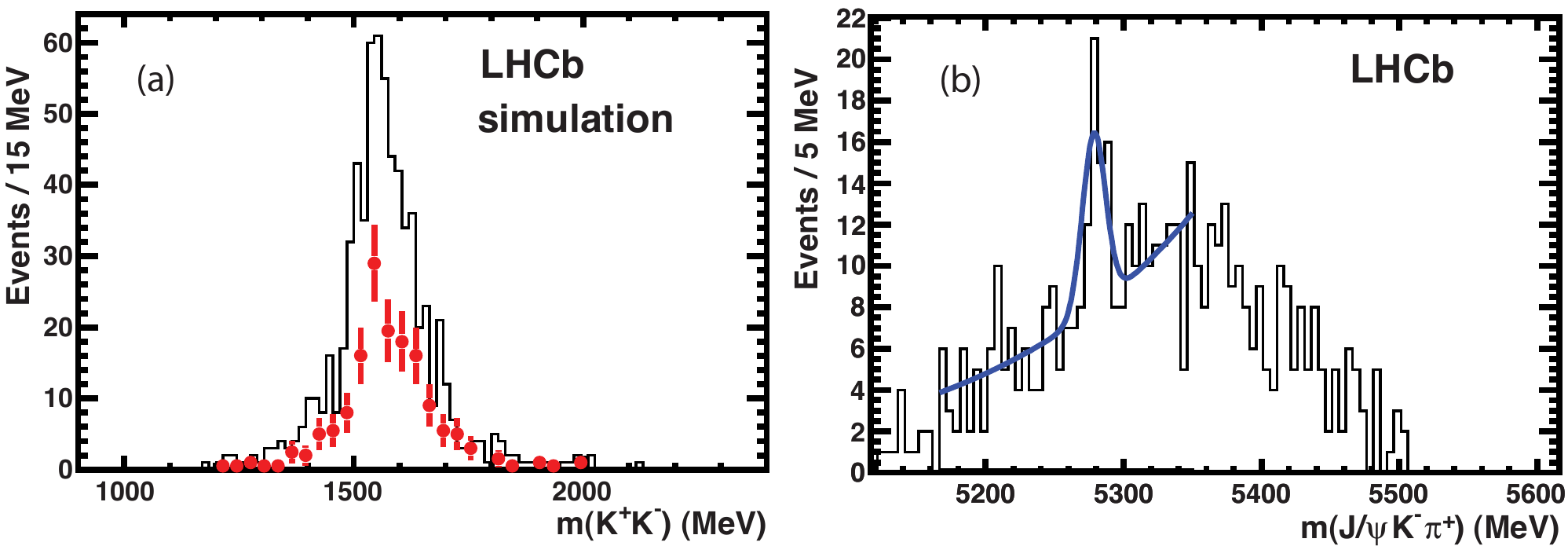}
\caption{(a) The $m(K^+K^-)$ distribution for simulated $\overline{B}^0\to J/\psi \overline{K}_2^*(1430)$ decays where the $\pi^+$ from the $\overline{K}_2^*(1430)$ decay is interpreted as a $K^+$.  The histogram shows  $m(K^+K^-)$ in the signal region of $\Bs$ mass and the points in the sideband region. The simulation corresponds to approximately 8 fb$^{-1}$ of data. (b) The $m(J/\psi K^+\pi^-)$ distribution for $J/\psi K^+K^-$ data candidates $25-200$\,MeV above the $\Bs$ mass, and with $m(K^+K^-)$ within $\pm$300 MeV of 1525 MeV, reinterpreted as $\overline{B}^0\to J/\psi K^-\pi^+$ events. The fit is to a signal Gaussian whose mass and width are allowed to vary as well as a quadratic background.} 
\label{MC-data}
\end{figure}

It is possible for the decay $\overline{B}^0\to J/\psi K^-\pi^+$ to fake our signal if the $\pi^+$ is misidentified as a $K^+$. A specific example is given by $\overline{B}^0\to J/\psi \overline{K}_2^*(1430)$ decays \cite{:2008nk}.
%Before claiming this is a new $\Bs$ decay mode we consider whether part or all of the signal is from other $B$ to $J/\psi$ exclusive decay modes. 
%The BaBar collaboration observes a  $\overline{K}_2^*(1430)$ signal in $\overline{B}^0\to J/\psi K^-\pi^+$ decays \cite{:2008nk}.
To examine if we are sensitive to a reflection of this mode in the 1525 MeV di-kaon  mass region, a simulation 
%of $\overline{B}^0\to J/\psi \overline{K}_2^*(1430)$ decays 
was performed where the $\pi^+$ from the $\overline{K}_2^*(1430)$ was interpreted as a $K^+$. Figure~\ref{MC-data}(a) shows that the reflection of this mode does indeed peak in the di-kaon mass range around 1525 MeV.
It also peaks in the $\Bs$ signal region but is much wider than the $\Bs$ mass peak.  The region $25-200$\,MeV above the $\Bs$ mass peak provides a sample of misidentified $\overline{B}^0\to J/\psi K^-\pi^+$ decays. By measuring the number of $\overline{B}^0$ candidates in this region we can calculate the number in the $\Bs$ signal region.

%No attempt was made to predict an absolute rate based on the $\pi$ to $K$ misidentification rate, and the relative rate of $B^0/B_s$ production \cite{fsfd}, because of the unmeasured branching ratio.

%Evidently we can be victims of a reflection here and the sideband subtraction does not subtract off enough events to remove the effect. Thus we need to ascertain the size of this effect. 
To determine the size of any $\overline{B}^0$ reflection in the $f'_2$ mass region
%, defined as being within $\pm$300 MeV of 1525 MeV, 
we select events where the reconstructed $J/\psi K^+K^-$ mass is in the range $25-200$\,MeV above the $\Bs$ mass, reassign each of the two kaons in turn to the pion hypothesis, and plot the $J/\psi K\pi$ mass.
The resulting peak at the $\overline{B}^0$ mass has
%has 594$\pm$39 events using loose kaon identification cuts, and 
36$\pm$10 events from the fit shown in Fig.~\ref{MC-data}(b).  We calculate 37$\pm$10 events in the 
$\Bs$ signal region, using the shape from Monte Carlo simulation, and use this number as a constraint in the fit described below to determine the $f'_2$ parameters and signal yields.

\begin{figure}[t!]
\centering
\includegraphics[width=6.in]{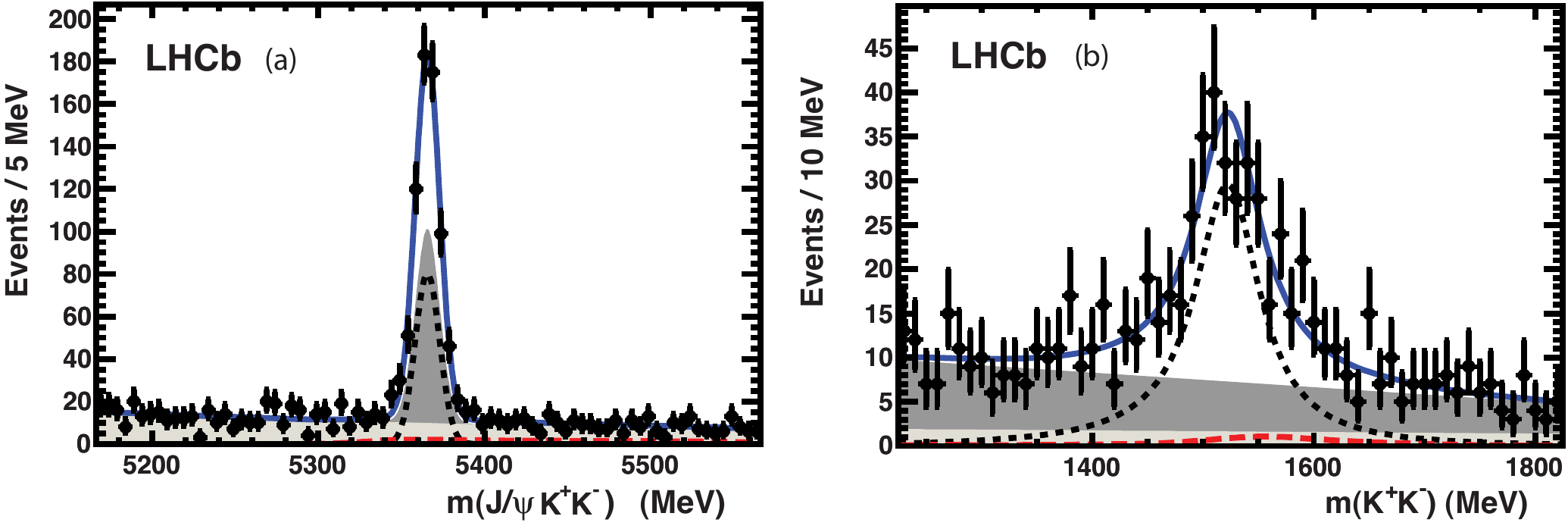}
\caption{Projections of fits to (a) the $\Bs$ candidate mass and (b) the di-kaon mass. The $f'_2$ signal is parameterized by a spin-2 Breit-Wigner function whose width is fixed to 73 MeV  (dotted curve). The combinatorial background is shown in the light shaded region, while the darker shaded region shows the non-resonant $J/\psi K^+K^-$ component. The long-dashed (red) line shows the misidentified $\overline{B}^0\to J/\psi K^-\pi^+$ decays, and the (blue) line the total.}
\label{fit_mbs-mkk-f2-tight}
\end{figure}

\begin{figure}[hbt]
\centering
\includegraphics[width=4in]{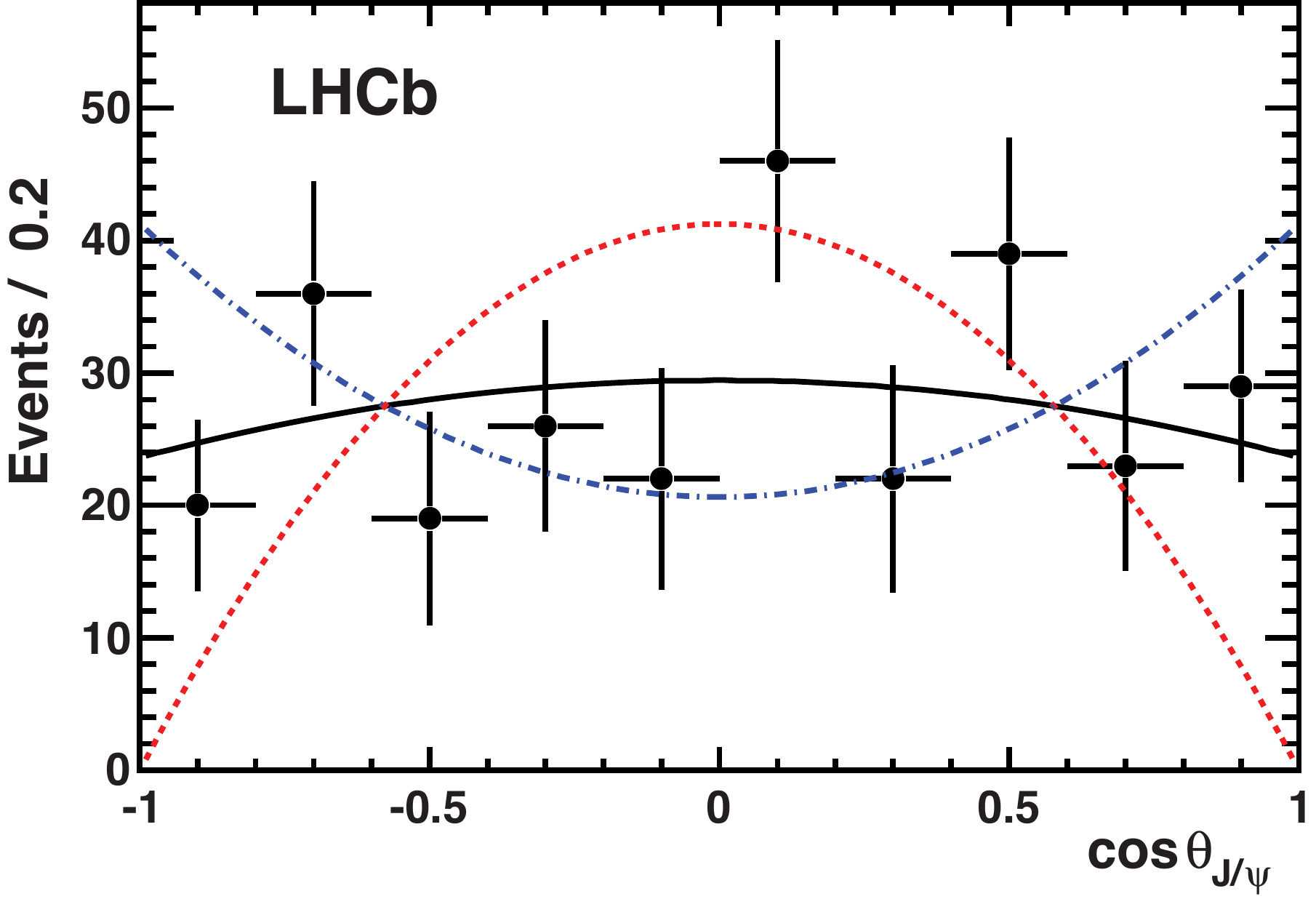}
\caption{Distribution of $\cos\theta_{J/\psi}$ for $\Bs\to J/\psi f'_2$ decays. The background and non-resonant $K^+K^-$ components have been subtracted, and the data have been corrected for acceptance. The  fit to Eq.~\ref{eq:jpsihel} is shown by the solid line. Note that for pure S-wave the distribution would be $\sin^2\theta_{J/\psi}$ ($p=0$), shown as the dotted curve, while for pure helicity 1 ($p=1$) the data would be described by the dot-dashed curve. } \label{heljpsi_f2}
\end{figure}

%Only tight kaon identification cuts are used subsequently for the di-kaon analysis.
To test the $f'_2$ hypothesis
we perform a simultaneous fit to the $\Bs$ candidate mass and the di-kaon mass. 
The $f'_2$ signal is parameterized by a spin-2 Breit-Wigner function \cite{Mizuk:2008me}.   The width of the $f'_2$ is fixed to the PDG value of 73\,MeV \cite{PDG}. 
 A contribution from non-resonant $K^+K^-$ is included as a linear function in the di-kaon mass. The contribution from the $K^-\pi^+$  reflection is included using the di-kaon and $\Bs$ mass shapes from the simulation, with the normalization fixed by the event yield in Fig~\ref{MC-data}(b).
The results of the fits are shown in Fig.~\ref{fit_mbs-mkk-f2-tight}. The $f_2'$ mass from the fit is 1525$\pm$4 MeV and the yield is 269$\pm$26 events within $\pm$125 MeV of the mass.  If we allow the $f_2'$ width to vary we find a consistent value of 90$^{+16}_{-14}$ MeV.
%Kaon ID &Mass (MeV) & Width (MeV) & $\overline{K}^{*0}$ Reflection& Signal events\\\hline
%Tight & 1532$\pm$5 & 90$^{+16}_{-14}$ &66$\pm$15 & 320$\pm$33 \\
%Tight & 1528$\pm$4 & 73 (fixed)& 65$\pm$15 & 296$\pm$26 \\
As we have not taken into account possible interferences between the $f'_2$ and other $J/\psi K^+K^-$ final states we do not provide systematic uncertainties for these values.
%The $f'_2(1525)$ has also been observed to decay into two pions but at a rate about 1\% of the di-kaon decay. The presence of the di-pion decay, however, does mean that the $G$-parity of this state is even and thus the spin must be even. 
The decay angle of the $J/\psi$, $\theta_{J/\psi}$,  can test for pure spin-0, or the presence of a higher spin state such as the spin-2 $f'_2$ \cite{PDG}. Here $\theta_{J/\psi}$ is defined as 
the angle of the $\mu^+$ with respect to the $\Bs$ direction in the $J/\psi$ rest frame. It is distributed as
\begin{equation}
f(\cos\theta_{J/\psi})=(1-p)\sin^2\theta_{J/\psi}+\frac{p}{2}\left(1+\cos^2\theta_{J/\psi}\right),
\label{eq:jpsihel}
\end{equation}
where $1-p$ is the fraction of helicity zero and $p$ is the fraction of helicity $\pm$1. 
Shown in Fig.~\ref{heljpsi_f2} is the background subtracted, acceptance corrected $\cos\theta_{J/\psi}$ distribution for $K^+K^-$ masses in the $f'_2$ region. MC simulation is used to find the acceptance correction. The points are extracted from the joint fit to the $m(J/\psi K^+K^-)$  and $m(K^+K^-)$ distributions in the $K^+K^-$ mass region within $1400-1650$\,MeV for events in the peak above the non-resonant $K^+K^-$.  The fit result is $p=(0.57\pm 0.13)$, with $\chi^2$/number of degrees of freedom (ndof) of 10/8 (27\% probability). Fitting only with an S-wave gives $\chi^2$/ndof of 27/9 (0.1\% probability),
showing that the data are not likely to be pure spin-0, but are compatible with a higher spin state consistent with an $f'_2$ contribution. 

%%\section{Relative yields of   $J/\psi f_0(980)$,  and $J/\psi f'_2(1525)$ to $J/\psi\phi$ events}
The branching fraction of $\Bs\to J/\psi f_2'$ relative to $\Bs\to J/\psi \phi$ is determined by assuming that the dominant background is S-wave and the signal D-wave, so there is no interference between them.\footnote{Although there can be interference as a function of the $K^+$ decay angle in the $f_2'$ rest frame, integrating over this variable causes the net result to be zero.}  The number of $J/\psi K^+K^-$ events is determined by a fit to the $\Bs$ mass distribution, within $\pm$20 MeV of the $\phi$ mass. A small S-wave component in the $\phi$ mass region of  (4.2$\pm$2.3)\% is subtracted  \cite{JpsiphiLHCb}. Although there are the same final state particles in both modes, the relative efficiency is (78$\pm$2)\%,  where the uncertainty arises from simulation statistics. The efficiency ratio differs from unity due to the different $p_{\rm T}$ distributions of the kaons in the final states.
The kaon identification efficiencies are corrected with respect to those given by the MC simulation using a sample of $D^{*+}$ decays, where the kaon can be selected without resorting to PID information. Typical corrections are on the order of 5\%.

To find the effective relative rate of $f'_2$ decays we use the fit where the width is allowed to vary.  There are 320$\pm$33 $f'_2$ events and 1774$\pm$42 $\phi$ events. 
Correcting for the relative efficiencies and the explicit branching fractions ${\cal{B}}\left(f'_2(1525)\to K^+K^-\right)=(44.4\pm 1.1)$\%, and 
${\cal{B}}\left(\phi\to K^+K^-\right)=(48.9\pm 0.5)$\% \cite{PDG}, we measure
\begin{equation}
R\equiv \frac{{\cal{B}}\left(\Bs\to J/\psi f'_2(1525)\right)}{{\cal{B}}\left(\Bs\to J/\psi \phi\right)}=(26.4\pm2.7\pm2.4)\%. 
\end{equation}

The systematic uncertainty on $R$ has several contributions, as listed in Table~\ref{tab:syserr}.
\begin{table}[!b]
\centering
\caption{Systematic uncertainties on $R$.}
\label{tab:syserr}
\begin{tabular}{lc}
\hline\hline
Source & Change (\%) \\\hline
$f'_2$ width & 6.3\\
Helicity & 4.0\\
Relative efficiency& 2.6\\
S-wave under $\phi$& 2.3\\
$K^+K^-$ mass dependent efficiency& 2.3 \\
Background shape & 1.3\\
$\Bs$ $p_{\rm T}$ distribution & 0.5 \\
$\Bs$ mass resolution & 0.5  \\
PID & 1.0  \\
%$f'_2(1525)$ width & 2.4\\
Signal shape & 1.0 \\
${\cal{B}}\left(f'_2(1525)\to K^+K^-\right)$ & 2.5\\
${\cal{B}}\left(\phi\to K^+K^-\right)$ &1.0\\
\hline
Total  & 9.2 \\
\hline
\end{tabular}
\end{table}
The largest source of uncertainty is $f'_2$ width. The error quoted reflects changing the width by one standard deviation from the fitted value of 90 MeV. The helicity amplitudes of the $J/\psi f'_2$ decay are unknown, unlike the $J/\psi\phi$ amplitudes which are well measured \cite{PDG}. The difference between the values obtained using helicity zero and helicity one $J/\psi$ MC samples is 4\% compared to our central value. The S-wave subtraction of the events in the $J/\psi\phi$ region causes a 2.3\% uncertainty.
We include an uncertainty for the efficiency as a function of $K^+K^-$ mass, as the tracking could be sensitive to the opening angle of the kaon pair.  Modifying the acceptance from a flat to linear function of mass changes the yield by 2.3\%.  Varying the $\Bs$ $p_{\rm T}$ distribution within limits imposed by the data results in a small 0.5\% change in the rate. Changing the mass resolution by its error results in a 0.5\% change.
 A PID uncertainty of 1\% is added to account for different momentum distributions of the kaons in the two final states. As a check we note that the ratio of the number of events in $J/\psi\phi$ with tight cuts to loose cuts on the kaon identification is (61$\pm$2)\% and the simulation gives a consistent (60$\pm$1)\%.
 % The resonance width variation within our determined error leads to a 2.4\% change in the yield.
  Variation of the background and signal shapes makes small differences.
%\section{Conclusions}

In conclusion, we have made the first investigation of the $\Bsb\to J/\psi K^+K^-$ final state over the entire range of $K^+K^-$ mass. There is a significant non-resonant component that extends under the $\phi$ region which can affect \CP violation measurements \cite{Stone:2008ak}. We have also
observed $\Bs\to J/\psi f'_2(1525)$ decays. 
The branching fraction ratio relative to $J/\psi\phi$ is 
\begin{equation}
\frac{{\cal{B}}\left(\Bs\to J/\psi f'_2(1525)\right)}{{\cal{B}}\left(\Bs\to J/\psi \phi\right)}=(26.4 \pm2.7\pm 2.4)\%, 
\end{equation}
assuming that the background does not interfere with the signal amplitude.
This decay mode can also be used to measure \CP violation in the $\Bs$ system, although a different transversity analysis than in $J/\psi\phi$ would be required as the final state is a combination of a spin-1 $J/\psi$ and a spin-2 $f'_2$ state. Some consideration has been given to measuring \CP violation in vector-tensor decays \cite{Sharma:2005ji}. 
%\section*{Acknowledgements}

We express our gratitude to our colleagues in the CERN accelerator
departments for the excellent performance of the LHC. We thank the
technical and administrative staff at CERN and at the LHCb institutes,
and acknowledge support from the National Agencies: CAPES, CNPq,
FAPERJ and FINEP (Brazil); CERN; NSFC (China); CNRS/IN2P3 (France);
BMBF, DFG, HGF and MPG (Germany); SFI (Ireland); INFN (Italy); FOM and
NWO (the Netherlands); SCSR (Poland); ANCS (Romania); MinES of Russia and
Rosatom (Russia); MICINN, XuntaGal and GENCAT (Spain); SNSF and SER
(Switzerland); NAS Ukraine (Ukraine); STFC (United Kingdom); NSF
(USA). We also acknowledge the support received from the ERC under FP7
and the Region Auvergne.

\clearpage
\newpage
\ifx\mcitethebibliography\mciteundefinedmacro
\PackageError{LHCb.bst}{mciteplus.sty has not been loaded}
{This bibstyle requires the use of the mciteplus package.}\fi
\providecommand{\href}[2]{#2}

%\bibliographystyle{LHCb}
%\bibliography{f0}

\end{document}